\documentclass[12pt]{article}
\usepackage{epsfig, amssymb}
\usepackage{graphicx,epsfig}
\usepackage{textpos}
\usepackage{rotating}
\usepackage[usenames,dvipsnames]{color}
\begin{document}
\begin{titlepage}
\thispagestyle{empty}
\begin{flushright}
\end{flushright}

\bigskip


\begin{center}
  \noindent{\Large \textbf
    {Stochastic quantization of conformally coupled scalar in AdS}}\\

\vspace{2cm} \noindent{Dileep
  P. Jatkar${}^{a}$\footnote{e-mail:dileep@hri.res.in} and Jae-Hyuk
  Oh${}^{b}$\footnote{e-mail:jack.jaehyuk.oh@gmail.com}}

\vspace{1cm}
{\it
Harish-Chandra Research Institute, Chhatnag Road, Jhunsi,
Allahabad-211019, India${}^{a}$\\[2mm] 
Department of Physics, Hanyang University, Seoul {\it 133-791},
Korea${}^{b}$\\ 
}
\end{center}

\vspace{0.3cm}
\begin{abstract}
  We explore the relation between stochastic quantization and
  holographic Wilsonian renormalization group flow further by studying
  conformally coupled scalar in $AdS_{d+1}$.  We establish one to one
  mapping between the radial flow of its double trace deformation and
  stochastic 2-point correlation function. This map is shown to be
  identical, up to a suitable field re-definition of the bulk scalar, to
  the original proposal in arXiv:1209.2242.
\end{abstract}
\end{titlepage}

\newpage

\tableofcontents
\section{Introduction and Summary}

The AdS/CFT correspondence relates $d$ dimensional field theory to
$d+1$ dimensional theory of gravity.  This relation has been explored
in great detail over the years in various context.  Stochastic
quantization\cite{Wu1,Paul1,Dijkgraaf:2009gr} is a formalism which studies non-equilibrium dynamics of
$d$ dimensional field theory which evolves along stochastic time
variable.  Resulting theory is interpreted as a $d+1$ dimensional
field theory.  There were proposals relating AdS/CFT corresponding to
the stochastic quantization in the past\cite{Lifschytz:2000bj,Polyakov:2000xc,Petkou1,Minic:2010pw}.  

Recently, we proposed a specific relation between AdS/CFT and
Stochastic quantization.  In \cite{Oh:2012bx}, we proposed that the Hamiltonian governing
the holographic Wilsonian renormalization equations\cite{Polchinski1,Hong1} in the AdS/CFT
correspondence is equal to the Fokker-Planck Hamiltonian of the
stochastic system.  It in turn implies the stochastic time is
identified with the radial variable in the AdS space.  We also showed
that our proposal works for theories which are invariant under Weyl
rescaling\footnote{In \cite{Oh:2012bx}, we have dealt with theories which are invariant under the scaling of
the background metric as $g_{\mu\nu} \rightarrow \lambda(r)g_{\mu\nu}$, where
$\mu$ and $\nu$ are spacetime indices and $\lambda(r)$ is an arbitrary radial coordinate $r$-dependent function.}.  
Using this relationship it was shown that the Stochastic
quantization correctly reproduces the radial evolution of the double trace coupling for the
boundary theory.  

This proposal is based on the direct analogy between the holographic
RG equation,
\begin{equation}
\partial_\epsilon \psi_H(\phi,r)=- \int_{r=\epsilon}d^dx \mathcal
H_{RG}(-\frac{\delta }{\delta \phi},\phi)\psi_H(\phi,r), 
\end{equation}
where, $\mathcal H_{RG}$ is Legendre transform of the bulk action in
AdS space, $\psi_H = e^{-S_B}$ and $S_B$ is the boundary effective
action and on the stochastic side, the Fokker-Planck equation
\begin{equation}
\partial_t \psi_S(\phi,t)=-\int d^d x \mathcal
H_{FP}(\frac{\delta}{\delta \phi},\phi)\psi_S(\phi,t), 
\end{equation}
where, $H_{FP}$ is the Fokker-Planck Hamiltonian, which can be derived
from the Fokker-Planck action by Legendre transform.  The stochastic
wave-functional is written in terms of the probability distribution
$P(\phi, t)$ and the classical action $S_c$ as
\begin{equation}
\psi_S(\phi,t)= P(\phi,t)e^{\frac{S_c(\phi(t))}{2}}.
\end{equation}

In fact, the relation between the boundary effective action
obtained by solving Hamilton-Jacobi equations derived from the bulk
action and stochastic 2-point correlator obtained from the solution of
Langevin equation addressed in \cite{Oh:2012bx} is given by
\begin{equation}
\langle \phi_p(t)\phi_{-p}(t)\rangle^{-1}_H=\langle \phi_p(t)\phi_{-p}(t)\rangle^{-1}_S-\frac{1}{2}\frac{\delta^2 S_c}{\delta \phi_p \delta\phi_{-p}},
\end{equation}
where $\langle \phi_p(t)\phi_{-p}(t)\rangle_S$ is stochastic 2-point correlation function, 
$\langle \phi_p(t)\phi_{-p}(t)\rangle^{-1}_H=\frac{\delta^2 S_B}{\delta
  \phi_p \delta \phi_{-p}}$ and the stochastic time `$t$' is identified to the radial coordinate `$r$' in AdS space . From the Fokker-Planck approach, it is
also shown that
\begin{equation}
S_B=\int^t_{t_0}dt^\prime d^dp\ \mathcal
L_{FP}(\phi(t^\prime),\partial\phi(t^\prime);t^\prime), 
\end{equation}
where $\mathcal L_{FP}$ is called Fokker-Planck Lagrangian
density. This relation with the boundary effective action is
consistent with (1.4).

In this paper we will extend our analysis to conformally coupled
scalar\footnote{Conformally coupled scalar theories have been
  discussed in the literature, especially in the AdS$_{4}$ context(see
  \cite{Sebastian2,Sebastian3}).}.  As we did earlier, namely cases
involving Weyl invariant theories, we will treat AdS metric as a fixed
background except that in this case we will consider conformal
coupling of the scalar field with spacetime scalar curvature.  Since
the background is maximally symmetric, conformal coupling terms shows
up in the action as a mass term for the scalar field.  Interestingly
this mass falls within the window above the Breitenlohner-Freedman
bound for any dimensional AdS space which allows alternative
quantization of the scalar field in the AdS
space\cite{Daniel1,Witten11,Klebanov:1999tb,Witten:2001ua,Ioannis1,Sebastian1,Jatkar:2012mm}. We
can therefore study double trace coupling obtained by carrying out
alternate quantization.  From the stochastic quantization point of
view this example poses a new problem.  The Langevin equation for this
system turns out to have explicit stochastic time dependence.
Nevertheless, as we will see, it is still possible to use the Langevin
equation to determine equal time two-point correlation function.  We
will also be able to extract the Fokker-Planck action by eliminating
the noise term using the Langevin equation.

It turns out that the above relations, proposed in \cite{Oh:2012bx},
are still valid provided the classical action $S_c$ is obtained in a
more general way, which is the crucial ingredient in the above
relation.  In fact, one should be careful in choosing the classical
action because in general there are divergences and one may need to
add counter terms to regulate them.  Similar issue
arises for the classical action $S_c$ in case of conformally coupled
scalar in AdS space.
In \cite{Oh:2012bx}, it was proposed that $S_c=-2I_{os}(\phi_0)$,
where {\bf (1)$I_{os}$ is bulk on-shell action computed on AdS
  boundary(at $r=0$, where $r$ is the radial coordinate of AdS
  space).  Moreover, (2)there as no need to add counter term action
  in case of examples discussed in \cite{Oh:2012bx}, because those
  examples involved Weyl invariant bulk actions only.  It turns
  out that Weyl invariant bulk actions do not give rise to divergent terms
  at the AdS boundary.\footnote{These issues are addressed in the
    conclusion section of \cite{Oh:2012bx}.}}

Conformally coupled scalar action does give rise to divergences near
AdS boundary since it is not exactly Weyl invariant theory even if
it does enjoy certain scaling properties.  Therefore the natural question
that arises is how do we deal with these divergences. Our prescription is
that {\bf the bulk on-shell action, $I_{os}(\phi(\epsilon))$ is
  obtained at a certain radial cut-off, $r=\epsilon$ without adding
  any counter terms}, where $\phi(\epsilon)$ is the boundary value of
the bulk scalar field at $r=\epsilon$ and then $I_{os}$ should be written in
terms of $\phi(\epsilon)$. {\bf The classical action $S_c$ is then
  defined using the same relation,
  $S_c(\phi(\epsilon))=-2I(\phi(\epsilon))$ but at the radial cut off.}

The new definition of the classical action makes sense since it
correctly reproduces the classical actions for Weyl invariant cases,
and so does the expected form of stochastic 2-point correlation
functions. The on-shell action depends on radial cut-off $r=\epsilon$
explicitly in general, and that can be translated to the explicit
stochastic time dependence of the classical action $S_c$ defined on a
certain time slice $t=\epsilon$ when the radial coordinate $r$ is
identified with $t$.

We will then show that same result can be derived in a more elegant
way by doing field redefinition,
\begin{equation}
  \label{eq:1}
  \phi(t, p) = \Omega(t)f_p(t),
\end{equation}
where $\Omega(t)$ is a certain stochastic time $t$-dependent
function\footnote{In fact, $\Omega(t)$ should be restricted by a
  certain differential equation so that using the field redefinition
  consistency between Langevin and Fokker-Planck approaches can be
  established.}.
Interesting feature of this field redefinition is that the Langevin
dynamics in terms of $f_p(t)$ does not contain explicit dependence on
the stochastic time.  In fact in terms of $f_p(t)$ the system becomes
quite similar to that studied in the Weyl invariant examples.  This
analysis gives result consistent with that obtained without doing the
field redefinition.  Thus while appropriate Langevin and Fokker-Planck
descriptions can be derived even when there is explicit stochastic
time dependence, we also can access conventional description by doing
a field redefinition.  In other words, we can retain essence of our
proposed relation between AdS/CFT and stochastic quantization if we
allow for field redefinition.

This paper is organized as follows, in section 2 we will discuss
holographic Wilsonian Renormalization Group description of conformally
coupled scalar in AdS$_{d+1}$.  We solve for double trace deformation
both for zero as well as non-zero momenta.  To draw analogy with the
field redefinition that is we will carry out while studying the
Langevin dynamics, we will study effect of field redefinition on the
AdS side.  In section 3, we study stochastic quantization by first
studying stochastic time dependent Langevin equation and the deriving
the Fokker-Planck action.  In section 4, we carry out the field
redefinition and show that in the new variable, both the Langevin as
well as the Fokker-Planck dynamics take canonical form and the
original dictionary relating stochastic quantization to AdS/CFT can be
applied without any modification.

\section{Holographic Wilsonian renormalization 
group(HWRG) for conformally coupled scalar 
in $AdS_{d+1}$}
\label{Holographic Wilsonian renormalization group(HWRG) for
  conformally coupled scalar in}
\setcounter{equation}{0}
In this section, we derive Hamilton-Jacobi equations for the
holographic Wilsonian RG and their solutions for conformally coupled
scalar in $AdS_{d+1}$.
\subsection{Conformally coupled scalar and the radial flow of its
  double trace deformations}
\label{Conformally coupled scalar and its double trace deformations}
We start with the full bulk action for a scalar field propagating in $AdS_{d+1}$ as
\begin{equation}
\label{scalar-bulk-action}
S=\int_{r>\epsilon}drd^d x \sqrt{g}\mathcal L(\phi,\partial \phi)+S_B,
\end{equation}
where $S_B$ is the boundary effective action and the bulk Lagrangian
density $\mathcal L$ is defined as
\begin{equation}
\label{bulk-l}
\mathcal L=\frac{1}{2}g^{\mu\nu}\partial_\mu \phi \partial_\nu 
\phi +\frac{1}{2}m^2\phi^2 + \frac{\lambda}{4}\phi^{\frac{2(d+1)}{d-1}},
\end{equation}
where $g_{\mu\nu}$ is Euclidean $AdS_{d+1}$ metric, which is given by
\begin{equation}
ds^2=g_{\mu\nu}dx^\mu dx^\nu=\frac{dr^2 + \sum_{i=1}^d dx^i dx^i}{r^2}.
\end{equation}
$g^{\mu\nu}$ is the inverse metric, $\mu,\nu...$ run from 1 to $d+1$
whereas $i,j...$ run from 1 to $d$. $\epsilon$ is an arbitrary radial
cut-off.  The higher order interaction term in (\ref{bulk-l}) is
rather ill-defined since the power of it will be fractional in
general. However, it is well defined in a certain bulk dimensions, for
example, it becomes $\phi^4$ interaction in $AdS_4$ and $\phi^3$ in
$AdS_6$ respectively\footnote{4-dimensional case is conformally
  coupled scalar in $AdS_4$. For detailed discussion, see
  \cite{Petkou1}}.  In what follows, we will choose
$m^2=-\frac{d^2-1}{4}$ and will set $\lambda=0$ to deal with free
theory for a moment.  We point out that there are two different merits
when the mass of the scalar field is chosen to be
$m^2=-\frac{d^2-1}{4}$. Firstly, this mass value is in the window of
mass square of the scalar field $-\frac{d^2}{4}\leq m^2 \leq
-\frac{d^2}{4}+1$. {\bf  In such a case, alternative quantization in the dual
CFT defined on the boundary of AdS space is possible}, and then we
have two different fixed points for the double trace deformation
coupling in $UV$ region. Secondly, it will show a scaling property
that will be discussed in the next subsection. This allows us to deal with
this theory from a different view point and provides a more rigorous way of
defining relation between SQ and HWRG of this theory.

As usual, in order to derive the Hamilton-Jacobi type
HWRG flow equation\cite{Polchinski1,Hong1}, we take derivative of the
bulk action(\ref{scalar-bulk-action}) with respect to $\epsilon$ (the
radial cutoff), and
impose the condition that the full bulk action $S$ does not depend on
the radial cut-off. The Hamilton-Jacobi equation thus obtained is given by
\begin{equation}
\partial_\epsilon S_B=-\int_{r=\epsilon} d^d x\left[ \frac{1}{\sqrt{g}g^{rr}}\left( \frac{\delta S_B}{\delta \phi(x)}\right) \left( \frac{\delta S_B}{\delta \phi(x)}\right)
-\sqrt{g}\mathcal L(\phi,\partial \phi) \right].
\end{equation}
It is convenient to solve the above equation in momentum space by
using the Fourier transform
\begin{equation}
\label{phi-fourior-transform}
\phi(x^\mu)=\frac{1}{(2\pi)^{d/2}}\int^\infty_{-\infty}d^dp e^{-ip_i x_i}\phi_p(r).
\end{equation}
The HWRG equation in the momentum space then becomes
\begin{equation}
\label{Hamilton-Jacobi-for-con-scalar}
\partial_\epsilon S_B=-\int_{r=\epsilon} d^d p\left[ \frac{1}{2\sqrt{g}g^{rr}}\left( \frac{\delta S_B}{\delta \phi_p}\right) \left( \frac{\delta S_B}{\delta \phi_{-p}}\right)
-\frac{1}{2}\sqrt{g}g^{ij}p_ip_j \phi_p \phi_{-p}+\frac{d^2-1}{8}\sqrt{g}\phi_p \phi_{-p}\right],
\end{equation}
where the AdS$_{d+1}$ metric $g^{ij}=r^2\delta_{ij}$ and $\delta_{ij}$
is the Kronecker delta function.  To solve this equation, we propose the
following form of the boundary effective action:
\begin{equation}
S_B=\Lambda(\epsilon)+\int \frac{d^d p}{(2\pi)^d} \sqrt{\gamma}\mathcal J(\epsilon,p)\phi_{-p}
-\int \frac{d^d p}{2(2\pi)^d}\sqrt{\gamma}\mathcal D(\epsilon,p)\phi_p\phi_{-p},
\end{equation}
where $\mathcal D$ is  the ``double-trace'' coupling, $\mathcal J$ is
the boundary source term and $\Lambda$ is the boundary cosmological
constant. Substituting this ansatz into
Eq.(\ref{Hamilton-Jacobi-for-con-scalar}) and comparing the coefficients
of expansion in the boundary fields $\phi_p$, we get the following
three equations
\begin{eqnarray}
\partial_\epsilon \Lambda(\epsilon)&=&-\frac{1}{2}\int\frac{d^dp}{(2\pi)^{2d}}\frac{1}{\sqrt{g}g^{rr}}J(\epsilon,-p)J(\epsilon,p), \\
\partial_\epsilon J(\epsilon,p)&=&\frac{1}{\sqrt{g}g^{rr}(2\pi)^d}J(\epsilon,-p)D(\epsilon,p), \\ 
{\rm \ and \ \ }\partial_\epsilon D(\epsilon,p)&=&\frac{1}{\sqrt{g}g^{rr}(2\pi)^d}D(\epsilon,p)D(\epsilon,-p)-(2\pi)^d\sqrt{g}\left(r^2\delta_{ij}p_ip_j-\frac{d^2-1}{4}\right),
\end{eqnarray}
where $J(\epsilon,p)\equiv\sqrt{\gamma}\mathcal J(\epsilon,p)$,
$D(\epsilon,p)\equiv\sqrt{\gamma}\mathcal D(\epsilon,p)$ and
$\gamma=\frac{g(\epsilon)}{g_{rr}(\epsilon)}$ is the induced metric on the
$r=\epsilon$ hyper-surface.

As demonstrated as in \cite{Hong1}, the solution of double trace
coupling, $D$ is given by
\begin{equation}
\label{SoL-HJ-EQ}
D(\epsilon,p)=-(2\pi)^d\frac{\Pi_\phi}{\phi},
\end{equation}
where 
\begin{equation}
\label{definition-of-canonical-momentum}
\Pi_{\phi}=\sqrt{g}g^{rr}\partial_r \phi=\frac{\delta S_B}{\delta \phi}
\end{equation}
is canonical momentum of $\phi$ and it satisfies 
\begin{equation}
\label{Hamilton-equation-of-Pi}
\partial_r \Pi_{\phi}=\sqrt{g}\left(r^2|p|^2-\frac{d^2-1}{4}\right)\phi_p,
\end{equation}
in the classical gravity limit of the bulk theory.

\paragraph{Double trace deformation: zero momentum solution}
To examine the double trace deformation term $\mathcal D$, we need to
solve bulk equations of motion for the conformally coupled scalar. The
bulk equation of motion is given by
\begin{equation}
\label{bulk-equation-of-motion}
0=g^{\mu\nu}\nabla_\mu\partial_\nu \phi(x)+\frac{d^2-1}{4}\phi(x)-\frac{\lambda(d+1)}{2(d-1)} \phi^{\frac{d+3}{d-1}},
\end{equation}
where $\nabla_\mu$ is covariant derivative.  In fact, this is also
given by combining Eq.(\ref{definition-of-canonical-momentum}) and
Eq.(\ref{Hamilton-equation-of-Pi}) in momentum space in the limit
$\lambda=0$,
\begin{equation}
0=\partial^2_r \phi_p-\frac{d-1}{r}\partial_r \phi_p +\left( \frac{d^2-1}{4r^2}-p^2 \right)\phi_p,
\end{equation}
where $p^2=\sum_{i,j=1}^d p_ip_j \delta_{ij}$. In the zero momentum
limit, $p_i=0$, the most general solution is given by
\begin{equation}
\phi=a_1r^{\frac{d-1}{2}}+a_2r^{\frac{d+1}{2}},
\end{equation}
where $a_1$ and $a_2$ are arbitrary constants. Using the solution of
Hamilton-Jacobi equation (\ref{SoL-HJ-EQ}), the double trace coupling
becomes
\begin{equation}
\mathcal D(r)=\frac{D(r)}{\sqrt{\gamma}}=-(2\pi)^d\frac{d-1}{2}\left( \frac{\frac{d+1}{d-1}r+\chi}{r+\chi} \right),
\end{equation}
where $\chi=\frac{a_1}{a_2}$.  There are two different fixed points
for the double trace coupling, $\mathcal D(r)$ at {\it UV} region,
$r=0$.  When $\chi=0$, the double trace coupling has $\mathcal
D(r=0)=-(2\pi)^d\left(\frac{d+1}{2}\right)$ at the {\it UV} region and
it is a fixed point. Another fixed point is obtained when
$\chi=\infty$. In this case, $\mathcal
D(r=0)=-(2\pi)^d\left(\frac{d-1}{2}\right)$.  In the {\it IR} region,
$r=\infty$, the fixed points exist. When $\chi=\infty$, $\mathcal
D(r=\infty)=-(2\pi)^d\left(\frac{d-1}{2}\right)$ is fixed point.  For
the other generic value of $\chi$ including $\chi=0$, $\mathcal
D(r=\infty)=-(2\pi)^d\left(\frac{d+1}{2}\right)$ is fixed point.

Finally, the double trace deformation part of boundary effective
action $S_B$ is given by
\begin{equation}
S^{DT}_B=\frac{1}{2}\left( \frac{d-1}{2r^d} \right)\left( \frac{\frac{d+1}{d-1}r+\chi}{r+\chi} \right)\phi^2.
\end{equation}

\paragraph{Solution with non-zero momenta}
The most general solution of this equation of motion with non-zero
momenta $p_i$ is
\begin{equation}
\label{the-most-general-sol-with-non-zero-momenta}
\phi_p=r^{\frac{d-1}{2}}\left[\phi_0(p) \cosh(|p|r)+\phi_1(p)\sinh(|p|r)\right],
\end{equation}
where $|p|$ is norm of $p_i$, $\phi_0(p)$ and $\phi_1(p)$ are
arbitrary momentum dependent functions.  Conjugate momentum
$\Pi_\phi(p)$ is
obtained using its definition (\ref{definition-of-canonical-momentum})
as
\begin{equation}
 \Pi_\phi(p)=\frac{\frac{d-1}{2}\phi_0(p)+|p|r\phi_1(p)}{r^{\frac{d+1}{2}}} \cosh(|p|r)+\frac{\frac{d-1}{2}\phi_1(p)+|p|r\phi_0(p)}{r^{\frac{d+1}{2}}}\sinh(|p|r).
\end{equation}
The double trace deformation coupling, $\mathcal D(r,p)$ is then given by
\begin{equation}
\mathcal D(r,p)=\frac{D(r,p)}{\sqrt{\gamma}}=-(2\pi)^d\left[\frac{d-1}{2}+|p|r\frac{ \sinh(|p|r)+\tilde \phi(p)\cosh(|p|r)}{\cosh(|p|r)+\tilde \phi(p)\sinh(|p|r)}\right],
\end{equation}
where $\tilde \phi(p)=\frac{\phi_1(p)}{\phi_0(p)}$.
Finally, the double trace part of the boundary effective action $S_B$
becomes
\begin{equation}
\label{phi-double-trace}
S_B
=-\frac{1}{2}\int \frac{d^d p}{(2\pi)^d} \frac{\mathcal D(r,p)}{r^d}
\phi_p\phi_{-p},
\end{equation}
where we have explicitly written down only the double trace
deformation term in $S_B$ and we will do the same for any $S_B$
appearing hereafter unless stated otherwise.
\subsection{Re-defined field and its relation with the original field $\phi$}
\label{Re-defined field and relations with its original field}
We start from the bulk action(\ref{scalar-bulk-action}) and define a
new field\footnote{The properties of this redefined field are
  discussed in \cite{Petkou1}.}  $f(x^\mu)$ which is related to 
the original field $\phi$ by a field redefinition,
\begin{equation}
\label{field-redefinition}
\phi(x^\mu)=\Omega(r)f(x^\mu),
\end{equation}
where we will choose $\Omega(r)\equiv r^{\frac{d-1}{2}}$. Using this
field re-definition, the bulk action(\ref{scalar-bulk-action}) can be
written as
\begin{equation}
\label{action-with-redefined-field}
S=\int_{r>\epsilon} dr d^dx \left(
  \frac{1}{2}\delta^{\mu\nu}\partial_\mu f(x) \partial_\nu f(x) 
+\frac{\lambda}{4}f^{\frac{2(d+1)}{d-1}}(x)\right)+\frac{d-1}{2}\int
d^dx\left.\frac{f^2(x)}{2r}\right|^\infty_\epsilon+S_B, 
\end{equation}
where we have used a relation that
$g_{\mu\nu}=r^{-2}\delta_{\mu\nu}$. Up to boundary terms (the 
second term in the action(\ref{action-with-redefined-field})), the
bulk action becomes effectively that of a massless scalar field, $f(x)$
defined in $d+1$-dimensional flat Euclidean spacetime with
$f^{\frac{2(d+1)}{d-1}}(x)$ interaction.  Varying this bulk action
with respect to $f(x)$ provides a bulk equation of motion as
\begin{equation}
\label{f-bulk-equation-of-motion}
0=\delta^{\mu\nu}\partial_\mu\partial_\nu
f(x)-\frac{\lambda(d+1)}{2(d-1)} f^{\frac{d+3}{d-1}}(x), 
\end{equation}
which, of course, reproduces Eq(\ref{bulk-equation-of-motion}) once we
substitute the field redefinition Eq.(\ref{field-redefinition}) into it.
An interesting observation is that once we define a new boundary
effective action as 
\begin{equation}
\label{SB-and-SB-prime}
S^{\prime}_B=S_B-\frac{d-1}{2}\int_{r=\epsilon} d^dx\frac{f^2(x)}{2r},
\end{equation}
then {\it``massive scalar with mass $m^2=-\frac{d^2-1}{4}$,
  $\phi^{\frac{2(d+1)}{d-1}}$ interaction and the boundary action
  $S_B$ in Euclidean $AdS_{d+1}$ becomes precisely the same with
  massless scalar field with $f^{\frac{2(d+1)}{d-1}}$ interaction
  defined in flat Euclidean upper half of the spacetime with the
  boundary term $S^\prime_B$ in the classical gravity limit''}.  In
the following discussion, we will set $\lambda=0$ so that we will be
dealing with free field $f(x)$.
\paragraph{Holographic Wilsonian renormalization group in terms of the field $f(x)$:}
Recalling in terms of the new field $f(x)$, our action is that of a free
massless field. As a result, our starting point is
\begin{equation}
\label{define-sb-prime}
S=\frac{1}{2}\int_{r>\epsilon} dr d^dx  \delta^{\mu\nu}\partial_\mu f(x) \partial_\nu f(x)  +S^\prime_B.
\end{equation}
The Hamilton-Jacobi equation in momentum space derived from this
action becomes
\begin{equation}
\label{HJ-f}
\partial_\epsilon S^\prime_B=-\frac{1}{2}\int_{r=\epsilon}d^dp\left[ \left( \frac{\delta S^\prime_B}{\delta f_p}\right) 
\left(\frac{\delta S^\prime_B}{\delta f_{-p}}\right)- |p|^2 f_p f_{-p} \right],
\end{equation}
and the ansatz of $S^\prime_B$ as
\begin{equation}
S^\prime_B=\Lambda^\prime(\epsilon)+\int \frac{d^d p}{(2\pi)^d} \sqrt{\gamma}\mathcal J^\prime(\epsilon,p)\phi_{-p}
-\int \frac{d^d p}{2(2\pi)^d}\sqrt{\gamma}\mathcal D^\prime(\epsilon,p)\phi_p\phi_{-p}.
\end{equation}
one can get an equation of the double trace coupling, $D^\prime(r,p)\equiv\sqrt{\gamma}\mathcal D^\prime(r,p)$ and its solution as
\begin{eqnarray}
\partial_\epsilon D^\prime(\epsilon,p)&=&\frac{1}{(2\pi)^d}D^\prime(\epsilon,p)D^\prime(\epsilon,-p)-(2\pi)^d|p|^2, \\
D^\prime(\epsilon,p)&=&-(2\pi)^d \frac{\Pi_f}{f_p},
\end{eqnarray}
where 
\begin{equation}
\label{f-momenta}
\Pi_f=\partial_r f_p=\frac{\delta S^{\prime}_B}{\delta f_{-p}},
\end{equation}
is the canonical momentum of the re-defined field, $f_{-p}$. 
Since equation of motion of $f_p(r)$ is given by
\begin{equation}
(\partial^2_r-|p|^2)f_{p}=0,
\end{equation}
its solutions are
\begin{eqnarray}
f_p&=&b_1+b_2r, {\rm \ \ for\ zero\ momentum\ case,\ }p_i=0, \\
&=&f_0(p) \cosh(|p|r)+f_1(p)\sinh(|p|r), {\rm \ \ for\ nonzero\
  momentum\ case,} 
\end{eqnarray}
where $b_1$, $b_2$, $f_0(p)$ and $f_1(p)$ are arbitrary constants but
the last two are momentum dependent. Properties of fixed points of the
double trace coupling have similar behavior as the massless scalar field
defined in 2-dimensional Euclidean space(See examples in
\cite{Oh:2012bx} for detailed discussion.), and we will not discuss it
here.  We just list the precise forms of the double trace part of the
boundary effective action for further discussion:
\begin{eqnarray}
S^\prime_B&=& \frac{1}{2}\left( \frac{\tilde b}{1+\tilde b r} \right)f^2, {\rm \ \ for\ zero\ momentum,\ }p_i=0, \\
\label{result-redefined-field-f}
&=&\frac{1}{2}\int  {|p|d^d p}\left( \frac{\sinh(|p|r)+\tilde f_p \cosh(|p|r)}{\cosh(|p|r)+\tilde f_p \sinh(|p|r)}\right)f_p f_{-p}, {\rm \ \ for\ nonzero\ momentum,}
\end{eqnarray}
where $\tilde b=\frac{b_2}{b_1}$ and $\tilde f_p=\frac{f_1(p)}{f_0(p)}$.

\subsection{Relations between the schemes with $\phi$ and $f$}
In this subsection, we will discuss the relation between holographic
Wilsonian renormalization groups of the primitive field $\phi_p(r)$
and the rescaled field $f_p(r)$.  As mentioned, the two fields are
related by $\phi_p(r)=\Omega(r)f_p(r)$ and it turns out that
Hamilton-Jacobi equations of the two fields,
(\ref{Hamilton-Jacobi-for-con-scalar}) and (\ref{HJ-f}) are also
clearly transformed from one to another\footnote{To cancel the
  irrelevant terms in derivation, it is useful to substitute explicit
  form of $\Omega(r)=r^{\frac{d-1}{2}}$.}.  In order to perform such a
transformation, we have used the definition of canonical momenta
(\ref{definition-of-canonical-momentum}), (\ref{f-momenta}) in both
scheme, and following useful relations,
\begin{eqnarray}
\label{transformation-relations}
\frac{\delta S^\prime_B(f_p)}{\delta f_{-p}}&=&\Omega(\epsilon)\frac{\delta S_B(\phi_p)}{\delta \phi_{-p}}-(d-1)\frac{f_p}{2\epsilon}, {\ \ }
\frac{\delta S_B(\phi_p)}{\delta \phi_{p}}=\frac{1}{\Omega(\epsilon)}\frac{\delta S_B(f_p)}{\delta f_{p}},  \\ \nonumber
{\rm \ \ and \ \ }\partial_\epsilon S_B(\phi_p)&=&\partial_\epsilon S_B(f_p)-\int_{r=\epsilon} d^dp\frac{\delta S_B(f_p)}{\delta f_p}\frac{\partial_r \Omega(r)}{\Omega(r)}f_{-p},
\end{eqnarray}
which are derived from (\ref{SB-and-SB-prime}). The first two
relations in (\ref{transformation-relations}) are obvious.  The last
one uses the chain rule of differentiation.  While the first term on
the RHS is the usual change from $\phi$ to $f$, second term on the RHS
depends on the rescaling involved in the field redefinition.  In the
second term we extracts the $\Omega$ dependent piece from $S_B$ (which
has explicit $r$ dependence) and write its contribution as the cut-off
is varied.  Taking this factors in to account correctly, one gets the
last relation in (\ref{transformation-relations}).

The main relation between the two schemes is manifestly the relation between each double trace deformation, namely (\ref{SB-and-SB-prime}).
It can be easily proved that the double trace deformation parts (\ref{phi-double-trace}) and (\ref{result-redefined-field-f}) 
in each scheme are related to each other via (\ref{SB-and-SB-prime})
by the field re-definition $\phi_p=\Omega(r)f_p$.

\section{Stochastic quantization }
\setcounter{equation}{0}
In this section, we will develop the Langevin dynamics and the Fokker-Planck
approach respectively to reproduce the radial flows of double trace
deformation in massive scalar field in AdS space.
\subsection{Langevin equation with explicit time dependence}
In this section, we will find the Langevin equation which allows us to
derive the stochastic 2-point correlation function which, in turn, is
in one to one correspondence with the boundary effective action
obtained in the previous section via the relation obtained in
\cite{Oh:2012bx}:
\begin{equation}
\label{lagevin-relation}
\langle \phi_p(t)\phi_q(t)\rangle^{-1}_H=\langle \phi_p(t)\phi_q(t)\rangle^{-1}_S-\frac{1}{2}\frac{\delta^2 S_c}{\delta \phi_p \delta\phi_{-p}}, 
\end{equation}
where $\langle \phi_p(t)\phi_q(t)\rangle_S$ is the stochastic two point correlation function, $\langle \phi_p(t)\phi_q(t)\rangle^{-1}_H=\frac{\delta^2 S_B}{\delta \phi_p(t) \delta \phi_q(t)}$ and
$S_c$ is called classical action, which will be defined soon.

The Langevin equation that we want to solve has the following form:
\begin{equation}
\label{time-dependent-L-eq}
\frac{1}{\Omega(t)}\frac{d \phi_p(t)}{dt}=-\frac{1}{\Omega(t)}\left( |p|-\frac{\partial_t \Omega(t)}{\Omega(t)} \right)\phi_p(t)+\eta(t,p),
\end{equation}
where $p_i$ are $d$-dimensional momenta and $\eta(t,p)$ is the stochastic noise satisfying
\footnote{As discussed in
  literature\cite{Paul1,Oh:2012bx}, 1 and 2-point functions are
  given by
\begin{equation}
\langle \eta_{p}(t)\rangle =0, {\ \ }\langle \eta_{p}(t)\eta_{p^\prime}(t^\prime)\rangle =\delta^d(p-p^\prime)\delta(t-t^\prime).
\end{equation}
Expectation values of odd number of insertions of $\eta$ vanishes
and any even number of insertions of it will be re-written as
summation of all possible products of pairs of two point functions of $\eta$.  }
\begin{equation}
\label{etaeta-corelations}
\langle \eta(t,p)\eta(t^\prime,p^\prime)\rangle =\delta(t-t^\prime)\delta^d(p-p^\prime).
\end{equation}
Unlike the usual Langevin equation, the equation
(\ref{time-dependent-L-eq}) has explicit time dependence appearing
through $\Omega(t)$.  Consistency with the Fokker-Planck approach
requires $\Omega(t)$ to satisfy following condition,
\begin{equation}
\label{Delta-eq}
\frac{d^2 \Delta(t) }{dt^2}=\left(\frac{d^2-1}{4}\right)\Delta^{\frac{d+3}{d-1}}(t),
\end{equation}
where $\Delta(t)\equiv\frac{1}{\Omega(t)}$(A related discussion will
appear in the next subsection).

Since there is explicit time dependence in the Langevin equation, we
cannot follow the usual method of stochastic quantization.  We will
therefore propose a more general concept for the classical action given by
\begin{equation}
\label{general-form-of-Sc}
S^\phi_c=\int_{\tilde t=t} d^dp\frac{1}{\Omega^2(\tilde t)}\left( |p|-\frac{\partial_{\tilde t} \Omega(\tilde t)}{\Omega(\tilde t)} \right)\phi_p(\tilde t)\phi_{-p}(\tilde t).
\end{equation}
This definition is a bit strange when compared with the usual
procedure of stochastic quantization.  Normally the classical action
in stochastic quantization has no explicit time dependence.  We will
interpret this classical action and the resulting Langevin equation in
the following manner.  We define the classical action at $\tilde t=t$
time slice.  At that time slice, the time dependent factor
$\Omega(\tilde t)$ becomes a number as $\Omega(t)$.  The Langevin
equation satisfied by this classical action at any given time slice
is 
\begin{equation}
\frac{d \phi_p(t)}{dt}=-\frac{1}{2}\Omega^2(t) \frac{\delta S^\phi_c(\phi,t)}{\delta \phi_{-p}}+\Omega(t)\eta(t,p).
\end{equation}
This is equivalent to the Langevin equation (\ref{time-dependent-L-eq}).

The most general form of the solution of Langevin equation (\ref{time-dependent-L-eq}) is
\begin{equation}
\phi_p(t)=\Omega(t)\int^t_{t_0} dt^\prime e^{-|p|(t-t^\prime)}\eta(t^\prime,p).
\end{equation}
Using the $\delta$- function correlations of $\langle \eta(t,p)\eta(t,p^\prime)\rangle$ (\ref{etaeta-corelations}), 2-point equal time correlation function of $\phi_p(t)$ is obtained as
\begin{equation}
\label{general-stochastic-correlator}
\langle \phi_p(t)\phi_{p^\prime}(t)\rangle_S=\Omega^2(t)\frac{\delta^d(p-p^\prime)}{2|p|}\left( 1-e^{2|p|(t_0-t)} \right).
\end{equation}

\paragraph{Relation between Langevin dynamics and massive scalar in $AdS_{d+1}$ with non-zero momenta:}
Let us go back to the bulk theory in AdS$_{d+1}$ of a scalar field
with mass $m^2=-\frac{d^2-1}{4}$.  The most general form of the bulk
solution with non-zero momentum is given in
(\ref{the-most-general-sol-with-non-zero-momenta}).  This solution
diverges in the interior\footnote{We note that one should impose
  regularity condition on the bulk solution to evaluate bulk on-shell
  action, $I_{os}$, however, when on compute HWRG by using the most
  general solution (\ref{the-most-general-sol-with-non-zero-momenta}),
  there is no such regularity issue at all.}  . To remove this
divergence, we impose a regularity condition on the solution at the
Poincare horizon.  This gives a condition
$\phi_0(p)+\phi_1(p)=0$. This condition forces the solution to decay
exponentially as it approaches $r=\infty$.  The regular solution is
then given by
\begin{equation}
\label{regular-p-sol}
\phi_p(r)=\phi_0(p)r^{\frac{d-1}{2}}e^{-|p|r}.
\end{equation}
Using bulk equations of motion, boundary on-shell action at radial
cut-off $r=\epsilon$ can be obtained as 
\begin{equation}
\label{boundary-oh-shell}
S=\frac{1}{2}\int_{r=\epsilon}  d^dp \sqrt{g}g^{rr} \phi_p(r)\partial_r\phi_{-p}(r).
\end{equation}
With substitution of regular solution (\ref{regular-p-sol}) and using
explicit expression of the background metric into
(\ref{boundary-oh-shell}), we get
\begin{eqnarray}
I_{os}(r=\epsilon)&=&-\frac{1}{2}\int_{r=\epsilon}d^dp{e^{-2|p|r}}\left( |p|-\frac{d-1}{2r} \right)\phi_0(p)\phi_{0}(-p), \\ \nonumber
&=&-\frac{1}{2}\int_{r=\epsilon} \frac{d^dp}{r^{d-1}}\left( |p|-\frac{d-1}{2r} \right)\phi_p(r)\phi_{-p}(r)
\end{eqnarray}
This boundary on-shell action is not yet regularized since there is a
divergent term in it, namely the second term in the parenthesis.  This
divergence occurs as we take $r \to 0$ limit.
However, it turns out that to capture the radial evolution of the
corresponding double trace deformation, we can choose our classical action as
\begin{equation}
\label{classical-ACTion}
S_c(\epsilon)= -2I_{os}(\epsilon),
\end{equation}
at the radial cut-off $r=\epsilon$. The prescription for stochastic
quantization with such classical action is that (since we will
identify the radial variable $r$ to stochastic time $t$)
$S_c(\epsilon)$ becomes classical action defined at $t=\epsilon$ time
slice. In fact, the classical action (\ref{general-form-of-Sc}) from
bulk on-shell action (\ref{classical-ACTion}) can be reproduced by
substituting $\Omega(t)=t^{\frac{d-1}{2}}$.

The stochastic 2-point correlator is known from (\ref{general-stochastic-correlator}), which is given by
\begin{equation}
\label{stochastic-specific-relation}
\langle \phi_p(t)\phi_{p^\prime}(t)\rangle_S=t^{d-1}\frac{\delta^d(p-p^\prime)}{2|p|}\left( 1-e^{2|p|(t_0-t)} \right).
\end{equation}
It is clear that (\ref{stochastic-specific-relation}) precisely
reproduce the radial flow of double trace deformation,
\begin{equation}
\langle \phi_p(r)\phi_{p^\prime}(r)\rangle^{-1}_H\equiv\frac{\delta^2 S_B}{\delta \phi_{p}(r)\delta\phi_{p^\prime}(r)}=
\frac{1}{r^{d-1}}\left[\frac{d-1}{2}+|p|r\frac{\sinh(|p|r)+\tilde\phi(p) \cosh(|p|r)}{\cosh(|p|r)+\tilde\phi(p)\sinh(|p|r)}\right]
\end{equation}
via the relation (\ref{lagevin-relation}), when `$r$' is
identified to `$t$' and the initial time \footnote{The initial
  stochastic time is chosen so as to match 2-point stochastic
  correlations with the double trace deformation. For more detailed
  manipulations, see \cite{Oh:2012bx}} in
(\ref{stochastic-specific-relation}) is chosen as
$t_0=-\frac{1}{|p|}\coth^{-1}[\tilde \phi(p)]$. Here, the new constant
$\tilde \phi(p)$ is $\tilde \phi(p)=\frac{\phi_1(p)}{\phi_0(p)}$.

\subsection{The Fokker-Planck approach}
Fokker-Planck action is not precisely of the usual form in this
case. In fact, it has deformation from its original form by time
dependent factor $\Omega(t)$. In this section, we will derive the
correct form of Fokker-Planck Lagrangian, show that it has the same
form with bulk Lagrangian, and the double trace deformation will be
correctly obtained via the relation proposed in \cite{Oh:2012bx} ,
\begin{equation}
\label{SB-direction}
S_B=\int^t_{t_0}dt^\prime d^dp\mathcal L_{FP}(\phi(t^\prime),\partial\phi(t^\prime);t^\prime).
\end{equation}

To derive the Fokker-Planck action, the stochastic partition function is the best starting point:
\begin{equation}
\label{Z}
Z=\int[D\eta]exp\left( -\frac{1}{2} \int^t_{t_0} \eta_p(t^\prime)\eta_{-p}(t^\prime)d^dpdt^\prime\right).
\end{equation}
We substitute the Langevin equation (\ref{time-dependent-L-eq}) into
the partition function (\ref{Z}) to replace $\eta$ by stochastic field $\phi_p(t)$.
Functional integral measure part will transform by the Jacobian factor,
\begin{equation}
 J\left( \frac{\delta \eta}{\delta \phi}\right)=
\exp\left[\frac{1}{4}\int^t_{t_0}dt^\prime d^dp \ \Omega^2(t^\prime)\frac{\delta^2 S_c(\phi,t^\prime)}{\delta \phi_p(t^\prime) \delta \phi_{-p}(t^\prime)} \right].
\end{equation}
The stochastic partition function is given by
\begin{equation}
Z=\int [D\phi]e^{-S}=\int [D\phi]\exp\left[ -\int^t_{t_0} dt^\prime\int d^dp L(\phi,\partial \phi,t^\prime)\right],
\end{equation}
where	
\begin{eqnarray}
\label{phi-Fokker-Planck-LDensity}
L(\phi,\partial \phi,t)&=&\frac{1}{2\Omega^2(t)}
\left[\left(\frac{\partial \phi_p(t)}{\partial t}\right)\left(\frac{\partial \phi_{-p}(t)}{\partial t}\right)
+\frac{1}{4}\left(\frac{\delta S_c(\phi,t)}{\delta \phi_p}\right)\left(\frac{\delta S_c(\phi,t)}{\delta \phi_{-p}}\right)\right. \\ \nonumber
&-&\left.\frac{1}{2}\Omega^4(t)
\frac{\delta^2 S_c(\phi,t)}{\delta \phi_p(t) \delta \phi_{-p}(t)}+\Omega^2(t)\left(\frac{\delta S_c(\phi,t)}{\delta \phi_{-p}}\right)
\left(\frac{\partial \phi_{-p}}{\partial t}\right)\right].
\end{eqnarray}
The first term on the second line in (\ref{phi-Fokker-Planck-LDensity}) does not depend on field $\phi$ and it becomes an overall constant in the partition function $Z$.
The last term in $L$ is not a total derivative since classical action
contains explicit time dependence. To deal with $L$ more clearly, we
plug in the explicit form of the classical action
(\ref{general-form-of-Sc}). If we now assume that $\Omega(t)$ satisfies
(\ref{Delta-eq}), then $L(\phi,\partial \phi,t)$ can be brought into the
following form\footnote{In the stochastic partition function, the usual form of the exponent is as (\ref{S-action}). See equation (3.81) in \cite{Paul1}.}
\begin{equation}
\label{S-action}
S=\int^t_{t_0}dt^\prime\int^{\infty}_{-\infty}d^d pL(\phi,\partial \phi;t^\prime)=\int^t_{t_0}dt^\prime\int^{\infty}_{-\infty}d^dp
\mathcal L_{FP}(\phi,\partial \phi;t^\prime) + \frac{1}{2}\int^t_{t_0}dt^\prime\partial_{t^\prime} S^\phi_c(\phi,t^\prime),
\end{equation}
where $\mathcal L_{FP}$ is the Fokker-Planck Lagrangian density, which is given by
\begin{equation}
\label{EX-fokker-planck}
\mathcal L_{FP}=\frac{1}{2}\Omega(t)^{-\frac{2(d+1)}{d-1}}\left[ 
\Omega(t)^{\frac{4}{d-1}}\left(\frac{\partial \phi_p}{\partial t}\right)\left(\frac{\partial \phi_{-p}}{\partial t}\right) 
+ \Omega(t)^{\frac{4}{d-1}} |p|^2\phi_p\phi_{-p}-\frac{d^2-1}{4}\phi_{p}\phi_{-p}\right].
\end{equation}

We point out that $\mathcal L_{FP}$ has the same form as that of the
bulk Lagrangian density \footnote{Once $t$ is identified to $r$, then
  $\sqrt{g}=\Omega(t)^{-\frac{2(d+1)}{d-1}}$,
  $g^{rr}=g^{ii}=\Omega(t)^{\frac{4}{d-1}}$ provided by
  $\Omega(t)=t^{\frac{d-1}{2}}$.}  (\ref{bulk-l}) with
$m^2=-\frac{d^2-1}{4}$ when $\Omega(t)=t^{\frac{d-1}{2}}$ and `$t$' is
identified to `$r$'. $\Omega(t)=t^{\frac{d-1}{2}}$ is the solution of
equation (\ref{Delta-eq}). Therefore, there is no contradiction with
the previous derivation of $\mathcal L_{FP}$.

Finally, we develop double trace part of the boundary effective action
$S_B$ using the prescription (\ref{SB-direction}). Since
(\ref{EX-fokker-planck}) is a free theory on a certain time dependent
background, it is enough to evaluate $\mathcal L_{FP}$ using its
classical solutions if one does not consider back reaction.  Equation
of motion derived from (\ref{EX-fokker-planck}) is given by
\begin{equation}
\label{t-eq-with-t}
0=\partial^2_t \phi_p -\frac{d-1}{t}\partial_t \phi_p+\left( \frac{d^2-1}{4t^2}-|p|^2\right)\phi_p,
\end{equation}
and its most general form of solution is
\begin{equation}
\label{t-equation}
\phi_p(t)=t^{\frac{d-1}{2}}[\Phi_0(p)\cosh(|p|t)+\Phi_1(p)\sinh(|p|t)],
\end{equation}
with arbitrary $d$-momenta, $p_i$ dependent functions: $\Phi_0(p)$ and
$\Phi_1(p)$.  When we manipulate $S_B$, we can bring one term to
be proportional to (\ref{t-eq-with-t}).  The remaining term then is a
total derivative and contributes only a
boundary term.  With this manipulation (\ref{SB-direction}) becomes
\begin{equation}
\label{boundary-sdtd}
S_B=\frac{1}{2}\left.\int d^dp\frac{1}{\Omega^2(\tilde t)}\phi_p(\tilde t)\partial_{\tilde t}\phi_{-p}(\tilde t)\right|^{\tilde t=t}_{\tilde t=t_0}.
\end{equation}
To evaluate the correct boundary effective action, we set two boundary conditions. (1)The initial time $t_0$ is set to be
\begin{equation}
\label{initial-conDTion}
t_0=-\frac{1}{|p|}\coth^{-1}\tilde \Phi(p),
\end{equation}
where $\tilde \Phi(p)=\frac{\Phi_1(p)}{\Phi_0(p)}$. At $\tilde t=t_0$, the solution (\ref{t-equation}) 
of the equation of motion (\ref{t-eq-with-t}) becomes zero, $\phi_p(t_0)=0$.\footnote{For detailed discussion about such choice of the initial time, see \cite{Oh:2012bx}.}
(2) At $\tilde t=t$, we want $\phi_p(\tilde t=t)=\phi_p(t)$. Therefore, it is requested that 
\begin{equation}
\label{a-new-solution}
 \phi_p(\tilde t)=\left(\frac{\tilde t^{\frac{d-1}{2}}[\cosh(|p|\tilde t)+\tilde \Phi(p)\sinh(|p|\tilde t)]}
{t^{\frac{d-1}{2}}[\cosh(|p|t)+\tilde \Phi(p)\sinh(|p|t)]}\right)\phi_p(t).
\end{equation}
Substituting (\ref{a-new-solution}) into (\ref{boundary-sdtd}) and applying the initial boundary condition (\ref{initial-conDTion}) on it, we get
\begin{equation}
\label{final-sdtd}
S_B=\frac{1}{2}\int d^dp \frac{1}{t^d}\left( \frac{d-1}{2}+ |p|t\frac{\sinh(|p|t)+\tilde \Phi(p)\cosh(|p|t)}{\cosh(|p|t)+\tilde \Phi(p)\sinh(|p|t)}\right).
\end{equation}
It is easy to see that (\ref{final-sdtd}) is precisely the same with
(\ref{phi-double-trace}) once stochastic time `$t$' is identified to
the radial variable `$r$' in AdS space and $\tilde \phi(p)=\tilde
\Phi(p)$.

\section{Toward a better-defined Langevin equation via field re-definition}
\setcounter{equation}{0}
 Even though the Langevin equation (\ref{time-dependent-L-eq}) does not
look like that of the usual form, it might be justified that
(\ref{time-dependent-L-eq}) is the correct formulation by the fact
that the usual form of Langevin equation can be derived from it by a
field re-definition
\begin{equation}
\label{field-redefinition-stochastic}
\phi(t,p)=\Omega(t)f_p(t),
\end{equation}
where $f_p(t)$ is a new stochastic field. It turns out that the new field $f_p(t)$ satisfies a new Langevin equation
\begin{equation}
\label{Langevin-ff}
\frac{d f_p(t)}{dt}=-|p|f_p(t)+\eta(t,p),
\end{equation}
which can be easily derived from (\ref{time-dependent-L-eq}) by using (\ref{field-redefinition-stochastic}).
The first term on the right hand side of (\ref{Langevin-ff}) can be written as
\begin{equation}
|p|f_p(t)=\frac{1}{2}\frac{\delta S_c(\phi)}{\delta \phi_{-p}},
\end{equation}
which implies the classical action can be written as
\begin{equation}
\label{clean-classical-action}
S^f_c=\int d^dp|p|f_pf_{-p}.
\end{equation}
This is precisely what the authors present in \cite{Oh:2012bx} for the theory of massless scalar field in  2-dimensional flat space. Langevin equation (\ref{Langevin-ff}) 
has no explicit time dependent factors in it 
and nor does the classical action
(\ref{clean-classical-action}). Therefore, usual rules of stochastic
quantization can be applied to this classical action without any modifications.
We point out that this justification of the time dependent stochastic
dynamics is very similar to that presented in \cite{Haas1}.

%

\subsection{Stochastic quantization of $f(x)$}
It turns out that Langevin equation (\ref{Langevin-ff}) together with the classical action (\ref{clean-classical-action}) captures
the radial evolution of double trace operator $S^\prime_B$ defined in (\ref{define-sb-prime}) in the limit of free field theory.
Euclidean action $S_c$ will be identified to $-2I_{os}$ as demonstrated in (\ref{classical-ACTion}).
Using the bulk equation of motion(\ref{f-bulk-equation-of-motion}) in
momentum space using the Fourier transform as Eq.(\ref{phi-fourior-transform}) with $\lambda=0$, its on-shell action at $r=\epsilon$ cut-off is given by
\begin{equation}
\label{f-on-shell-action}
I_{os}=\frac{1}{2}\int_{r=\epsilon} d^d p  f_p(r) \partial_r f_{-p}(r).
\end{equation}
The bulk equation of motion in the momentum space is
\begin{equation}
\partial^2_r f_p(r)-p^2 f_p(r)=0,
\end{equation}
and the most general form of the solution is given by
\begin{equation}
f_p(r)=f_0(p)\cosh(|p|r)+f_1(p)\sinh(|p|r),
\end{equation}
where $f_0(p)$(the boundary value of the bulk field $f(x)$) and $f_1(p)$ are $r$-independent constants.
This solution should be regular in the interior of AdS space as
$r\rightarrow\infty$. To prevent divergent behavior of the solution,
we impose a condition
$f_0(p)+f_1(p)=0$.  Final form of the regular solution after imposing
the regularity condition is
\begin{equation}
\label{regular-f-solution}
f_p(r)=f_0(p)e^{-|p|r}.
\end{equation}
By using the explicit form of bulk solution (\ref{regular-f-solution}), we get
\begin{equation}
I_{os}=-\frac{1}{2} \int_{r=\epsilon} d^d p |p|f_p(r) f_{-p}(r)
\end{equation}

\paragraph{The Langevin dynamics $\&$ the Fokker-Planck approach:}
To evaluate stochastic 2-point correlator, we follow the prescription
given in \cite{Oh:2012bx}.  The Euclidean action is given by
\begin{equation}
S^f_c=-2I_{os}= \int d^d p |p|f_p(t)f_{-p}(t),
\end{equation}
where we identify the radial cut-off $\epsilon$ with the time slice $t$.
We plug the Euclidean action into Langevin equation 
\begin{equation}
\frac{df_p(t)}{dt}=-\frac{1}{2}\frac{\delta S^f_c}{\delta f_{-p}(t)}+\eta(p,t)=-|p|f_p(t)+\eta(p,t),
\end{equation}
where $\eta(p,t)$ is called the stochastic white noise which provides
interactions with the surroundings and has its 2-point
correlations as given in (\ref{etaeta-corelations}). The most general
solution of the Langevin equation then becomes
\begin{equation}
f_p(t)=\int^t_{t_0}d \tilde t e^{-|p|(t-\tilde t)}\eta(p,\tilde t),
\end{equation}
Choice of the initial time $t_0$ is obtained by following the
prescription given in \cite{Oh:2012bx},
\begin{equation}
t_0=-\frac{1}{|p|}\coth^{-1}(\bar f_p),
\end{equation}
where $\bar f_p$ an arbitrary momentum dependent function, which
should be chosen as $\bar f_p= \tilde f_p$ to reproduce the correct
double trace deformation, $S^\prime_B$ for the theory defined in
(\ref{define-sb-prime}).

For the final step, we evaluate 2-point correlator using correlation
functions of the stochastic noise, which is
given by
\begin{equation}
\label{ff-two-point-function}
\langle f_0(p)f_0(p^\prime)\rangle_S=\frac{1}{2|p|}\delta^d(p-p^\prime)\left( 1-\frac{\bar f_p -1}{\bar f_p +1}e^{-2|p|t} \right)
\end{equation}
It turns out that this stochastic 2-point function reproduces the
kernel of $S^\prime_B$, $\langle f_p f_{p^\prime}\rangle
^{-1}_H\equiv\frac{\delta^2 S^\prime_B}{\delta f_p \delta
  f_{p^\prime}}$ correctly through relation (\ref{lagevin-relation})
using $\frac{1}{2}\frac{\delta^2 S_c}{\delta f_p \delta
  f_{p^\prime}}=|p|\delta^d(p-p^\prime)$, when $r=t$ and $\bar f_p=
\tilde f_p$.  Fokker-Planck approach gives result which is consistent
with the Langevin dynamics.

\subsection{Transformation of the Fokker-Planck action with field re-scaling}
It is rather trivial that the Langevin equation with the original
field $\phi_p$ transforms into that with the rescaled field $f_p$
using the field re-definition(\ref{field-redefinition-stochastic}).
The new Langevin equation gives the consistent relationship between
the radial flow of the double trace deformation of massless scalar
field theory in flat space-time and the corresponding stochastic
quantization with the classical action(\ref{clean-classical-action})
as demonstrated in the last section.  In this section, to explain our
framework more clearly, we will demonstrate that the scale
transformation maps the time dependent Fokker-Planck action to the new
one without explicit time dependence and usual flat space form.  Let
us start with the action $S$ defined in (\ref{S-action}). The action
$S$ is comprised of two pieces: Fokker-Planck action and the total
derivative term with respect to $t$.  The total derivative term has
the form of $\int dt\partial_t S^\phi_c$, where $S^\phi_c$ is the
classical action defined in (\ref{general-form-of-Sc}). This is the
usual form of the action $S$ derived from the stochastic partition
function (\ref{Z}) \footnote{e.g. See \cite{Paul1,Dijkgraaf:2009gr}}. Now
what we want to show is that using the field rescaling
(\ref{field-redefinition-stochastic}), the action $S$ will transform
into the form of
\begin{equation}
S=S_{FP}(f_p)+\frac{1}{2}\int^t_{t_0} dt^\prime\partial_{t^\prime} S^f_{c},
\end{equation}
where $S_{FP}$ is the Fokker-Planck action in terms of the rescaled
field $f_p$ and $S^f_c$ is the classical action given in
(\ref{clean-classical-action}).

Once the relation (\ref{field-redefinition-stochastic}) is plugged into the action $S=\int dt \int d^d pL$ defined in (\ref{S-action}), it becomes
\begin{eqnarray}
\label{rescaled-L(f)}
L(f,\partial f;t)&=&\left[\frac{1}{2} \partial_tf_p(t) \partial_tf_{-p}(t) +\frac{1}{2}|p|^2 f_p(t) f_{-p}(t) \right] \\ \nonumber
&+&\frac{1}{2} f_p(t) f_{-p}(t)\left[ \frac{\Omega^{\prime 2}(t)}{\Omega^2(t)} -\frac{d^2-1}{4}\Omega^{-\frac{4}{d-1}}(t)\right]
+\frac{\Omega^{\prime}(t)}{\Omega(t)}\partial_t[f_p(t) f_{-p}(t)] \\ \nonumber
&+&\partial_t\left[\frac{1}{2}|p|f_p(t) f_{-p}(t) -\frac{\Omega^{\prime}(t)}{\Omega(t)}f_p(t) f_{-p}(t)\right].
\end{eqnarray} 
We point out that the scale factor $\Omega(t)$ is not arbitrary but it
is what satisfies the differential
equation (\ref{Delta-eq}). In terms of $\Omega(t)$, it becomes
\begin{equation}
\label{omega-relation}
\frac{\Omega^{\prime\prime}(t)}{\Omega(t)}-2\frac{\Omega^{\prime2}(t)}{\Omega^2(t)}=-\frac{d^2-1}{4}\Omega^{-\frac{4}{d-1}}(t).
\end{equation}
Using (\ref{omega-relation}), the term proportional to
$-\frac{d^2-1}{4}\Omega^{-\frac{4}{d-1}}(t)$ in the second line in
(\ref{rescaled-L(f)}) can be replaced by the left hand side of
(\ref{omega-relation}). Then, the second line in (\ref{rescaled-L(f)})
becomes total derivative and which precisely cancels the last term in
(\ref{rescaled-L(f)}). Finally, (\ref{rescaled-L(f)}) becomes
\begin{equation}
L(f,\partial f;t)=\left[\frac{1}{2} \partial_tf_p(t) \partial_tf_{-p}(t) +\frac{1}{2}|p|^2 f_p(t) f_{-p}(t) \right]+\partial_t\left[\frac{1}{2}|p|f_p(t) f_{-p}(t)\right].
\end{equation}
The terms in the first square bracket are precisely the Fokker-Planck
action and the term in total derivative is half of the
classical action (\ref{clean-classical-action}).  Therefore, the
Fokker-Planck actions in both schemes are clearly related by the scale
transformation.

\subsection{Relations between two different schemes of stochastic quantization with $\phi$ and $f$}
In both schemes with the original field $\phi$ and the new field $f$,
they satisfy the relations between their 2-point stochastic
correlation functions and double trace couplings in AdS/CFT
respectively. Namely, the theories with field $\phi$ satisfies the relation (\ref{lagevin-relation}) and for the new field $f_p$, the similar relation as
\begin{equation}
\label{the-new-field-the-relation}
\langle f_p(t)f_q(t)\rangle^{-1}_H=\langle f_p(t)f_q(t)\rangle^{-1}_S-\frac{1}{2}\frac{\delta^2 S^f_c}{\delta f_p \delta f_{-p}}
\end{equation}
is satisfied.

In fact, stochastic 2-point correlator in each scheme enjoy the relation as
\begin{equation}
\label{phi-f-stochastic-relation}
 \langle \phi_p(t)\phi_{-p}(t)\rangle_S=\Omega^2(t)\langle f_p(t)f_{-p}(t)\rangle_S.
\end{equation}
This is clear from (\ref{stochastic-specific-relation}) and (\ref{ff-two-point-function}). The classical actions in both theories also have a relation as
\begin{equation}
\label{phi-f-relation}
S^\phi_c(\phi)=S^f_c(f)-\int d^d p \frac{\partial_t \Omega(t)}{\Omega(t)}f_pf_{-p}.
\end{equation}
This relation is also well understood by looking at (\ref{Delta-eq}), (\ref{clean-classical-action}) and (\ref{field-redefinition-stochastic}). (\ref{phi-f-relation}) leads
to 
\begin{equation}
\label{phi-f-delta-relation}
\frac{\delta^2 S^f_c(f)}{\delta f_p \delta f_{-p}}=\Omega^2(t)\frac{\delta^2 S^\phi_c(\phi)}{\delta \phi_p \delta \phi_{-p}}+2\frac{\partial_t \Omega(t)}{\Omega(t)},
\end{equation}
where we have used $\frac{\delta}{\delta f_p}=\Omega(t)\frac{\delta}{\delta \phi_p}$. Using (\ref{phi-f-stochastic-relation}) and (\ref{phi-f-delta-relation}), one can 
manipulate the right hand side of (\ref{the-new-field-the-relation}) and obtain the relation between double trace deformations in the two different schemes.
Then, (\ref{the-new-field-the-relation}) becomes
\begin{equation}
\label{H-relation-f-and-phi}
\langle f_p(t)f_{-p}(t)\rangle ^{-1}_H
=\Omega^2(t)\left(\langle \phi_p(t)\phi_{-p}(t)\rangle^{-1}_H -\frac{\partial_t \Omega(t)}{\Omega^3(t)} \right),
\end{equation}
where we have used (\ref{lagevin-relation}) to switch the stochastic 2-point function with the double trace deformation in theory with the old field $\phi_p$.
This relation is precisely the same with the relation(\ref{SB-and-SB-prime}) between two different boundary effective actions, $S_B$ and $S^\prime_B$ obtained 
as the solutions of their Hamilton-Jacobi equations. It is clear that one can derive (\ref{H-relation-f-and-phi}) from (\ref{SB-and-SB-prime}) using definitions of
the double trace couplings as $\langle f_p(t)f_{-p}(t)\rangle^{-1}_H=\frac{\delta^2 S^\prime_B(f)}{\delta f_p \delta f_{-p}}$ 
and $\langle \phi_p(t)\phi_{-p}(t)\rangle^{-1}_H=\frac{\delta^2 S_B(\phi)}{\delta \phi_p \delta \phi_{-p}}$.

In summary,we have shown that all the rescaling arguments in the bulk
theories with scalar field with the specific mass square
$m^2=-\frac{d^2-1}{4}$ are consistent with their description with
stochastic quantization, in which one can also have scaling argument
and all the quantities are in one to one correspondence with
those quantities in the holographic description.

\section{Conclusion}
In this paper, we have constructed a precise one to one mapping
between holographic Wilsonian renormalization group(HWRG) of
conformally coupled scalar field in AdS$_{d+1}$ and stochastic
quantization(SQ) obtained from the classical action by identifying it
with the on-shell action of the bulk scalar field theory evaluated at
a certain radial cut-off of AdS space. Our Langevin equation and
Fokker-Planck Hamiltonian dynamics present explicit stochastic time
dependences in them and they cannot be dealt with the usual
methodology of SQ. However, we have suggested more general definition
of classical action and it turns out that SQ with such classical
action reproduces the radial evolution of the boundary effective
action of the conformally coupled scalar obtained from its HWRG
computation correctly. Moreover, we have proved that SQ with such
general definition of the classical action is consistent with the
usual stochastic quantization method up to a field redefinition.

This field re-scaling argument continues to be valid even when the theory
contains a certain class of interaction of the field $\phi$ of the type
$L_{int}\sim\lambda \phi^{\frac{2(d+1)}{d-1}}$.  Thus, it opens a new
playground where one investigates HWRG and SQ of interacting theories
and their mathematical relation.  The scaling property seems to
be very crucial ingredient to construct exact mapping between the two
schemes.

\section*{Acknowledgement}
Work of D.P.J. is partly supported by the the project
12-R$\&$D-HRI-5.02-0303 J.-H.Oh would like to thank his $\mathcal
W.J.$  Work of J.-H.Oh is supported by the research fund of Hanyang
University (HY-2013).

\end{document}